\documentclass[conference]{IEEEtran}
\usepackage[T1]{fontenc}

\usepackage{amssymb,amsmath,amsfonts,amsthm}
\usepackage{mathrsfs}
\usepackage{cite}
\usepackage{array}
\usepackage{tabularx}
\usepackage[table]{xcolor}
\usepackage{color}
\usepackage{mathtools}
\usepackage{subfigure}
\usepackage{mathtools}
\usepackage{tikz,pgf}
\usetikzlibrary{calc,positioning,mindmap,trees,decorations.pathreplacing}
\usepackage{graphicx}
\usepackage{bbm}
\usepackage[utf8]{inputenc}
\usepackage{algorithm}

\newcommand\irregularline[2]{%
	let \n1 = {rand*(#1)} in
	+(0,\n1)
	\foreach \a in {0.1,0.125,...,#2}{
		let \n1 = {rand*(#1)} in
		-- +(\a,\n1)
	} 
}

\IEEEoverridecommandlockouts
\ifCLASSINFOpdf
\else
\fi
\usepackage[cmintegrals]{newtxmath}
\begin{document}
\title{ Age of Information in a Multiple Access Channel with Heterogeneous Traffic and an Energy Harvesting Node}
	
	\author{
		\IEEEauthorblockN{Zheng~Chen\IEEEauthorrefmark{1}, Nikolaos~Pappas\IEEEauthorrefmark{2}, Emil~Bj\"{o}rnson\IEEEauthorrefmark{1}, and Erik~G.~Larsson\IEEEauthorrefmark{1}}
		\IEEEauthorblockA{\IEEEauthorrefmark{1}Dept. of Electrical Engineering, Link\"{o}ping University, Link\"{o}ping, Sweden.}\IEEEauthorblockA{\IEEEauthorrefmark{2}Dept. of Science and Technology, Link\"{o}ping University, Norrk\"{o}ping Campus, Sweden.\\ Email: \{zheng.chen, nikolaos.pappas, emil.bjornson@liu.se, erik.g.larsson\}@liu.se}
	\thanks{This work was supported in part by ELLIIT, CENIIT, and the Swedish Foundation for Strategic Research.}}
	\maketitle

\begin{abstract}	
	Age of Information (AoI) is a newly appeared concept and metric to characterize the freshness of data. In this work, we study the delay and AoI in a multiple access channel (MAC) with two source nodes transmitting different types of data to a common destination. The first node is grid-connected and its data packets arrive in a bursty manner, and at each time slot it transmits one packet with some probability. Another energy harvesting (EH) sensor node generates a new status update with a certain probability whenever it is charged. We derive the delay of the grid-connected node and the AoI of the EH sensor as functions of different parameters in the system. The results show that the mutual interference has a non-trivial impact on the delay and age performance of the two nodes. 
\end{abstract}

\begin{IEEEkeywords}
Age-of-Information, Energy Harvesting, Random Access, Queueing, Performance Analysis.
\end{IEEEkeywords}

\IEEEpeerreviewmaketitle

\section{Introduction}
The Age of Information (AoI) has attracted increasing attention as a new metric and tool to capture the timeliness of reception and freshness of data \cite{kosta2017age}. This concept first appeared in \cite{Altman2010, KaulSECON2011, KaulINFOCOM2012}. 
 
Consider a monitored source node which generates timestamped status updates, and transmits them through a wireless channel or through a network to a destination. The age of information that the destination has for the source, or more simply the AoI, is the elapsed time from the generation of the last received status update. Keeping the average AoI small corresponds to having fresh information. This notion has been extended to other metrics such as the value of information, cost of update delay, and non-linear AoI \cite{nonlinear_kosta, Sun2018sampling}.
The deployment of energy harvesting (EH) sensor networks is becoming an important aspect of the future wireless networks, especially in the Internet of Things (IoT) networks where devices opportunistically transmit small amounts of data with low power. Sensors with EH capabilities can convert ambient energy (e.g., solar power, thermal energy, etc.) into electrical energy, which allows for green and self-sustainable communication.

\subsection{Related Works}
Recently, several works have considered the AoI analysis and optimization in a network with an EH source node.
In \cite{BacinoglouITA2015}, the authors consider the problem of optimizing the process of sending updates from an EH source to a receiver to minimize the time average age of updates. Similar system and analysis can be found in \cite{lazy_timely, ArafaAsilomar2017, ArafaGC2017, age_eh2, age_eh3, ArafaICC2018, ArafaITA2018, age_eh1}. In \cite{BakninaCISS2018}, an erasure channel is considered between the EH-enabled transmitter and the destination. The transmitter sends coded status updates to the receiver in order to minimize the AoI. In \cite{BakninaISIT2018}, the authors consider the scenario where the timings of the status updates also carry an independent message. This information is transmitted through a receiver with EH capabilities and the trade-off between AoI and the information rate is studied.

The age-energy tradeoff is explored in \cite{BacinogluISIT2018}, where a finite-battery source is charged intermittently by Poisson energy arrivals. In \cite{FengINFOCOMWKSHPS2018}, the optimal status updating policy for an EH source with a noisy channel was investigated. In \cite{FengISIT2018}, the authors consider the case with update failures as the updates can be corrupted by noise. An optimal online updating policy is proposed to minimize the average AoI, subject to an energy causality constraint at the sensor.

Except the case with EH nodes harvesting ambient energy, some other works have considered wireless power transfer (WPT) to convert the received radio frequency signals to electric power \cite{ChenICCW2017}. 
In \cite{Krikidis2018}, the performance of a WPT-powered sensor network in terms of the average AoI was studied. The work in \cite{Abd-Elmagid2018} considers freshness-aware IoT networks with EH-enabled IoT devices. More specifically, the optimal sampling policy for IoT devices that minimizes the long-term weighted sum-AoI is investigated.

Since the status updates and regular information packets are associated with difference performance metrics, e.g., one with smallest AoI and the other with smallest delay, the impact of heterogeneous traffic on the AoI and the optimal update policy has been investigated in \cite{Stamatakis2018, KostaGC2018}. The work in \cite{game_coexistence} investigates Nash and Stackelberg equilibrium strategies for DSRC and WiFi coexisting networks, where DSRC and WiFi nodes are age and throughput oriented, respectively.

\subsection{Contribution}
Most of the existing works on AoI in EH-enabled networks consider a single transmitting node. When there is more than one node in the network with heterogeneous traffic and different types of power supplies, the effect of random data and energy arrivals on the performance of a multiple access channel (MAC) has not been studied.

In this work, we study the MAC where one node connected to the power grid has bursty arrivals of regular data packets, and another EH sensor sends status update when its battery is non-empty.  
We derive the data delay of the throughput-oriented node and the AoI of the EH sensor, which are given as functions of their transmission probabilities, the data arrival rate at the source node and the energy arrival rate at the sensor, which can be further used to optimize the operating parameters of such systems.

\section{System Model}

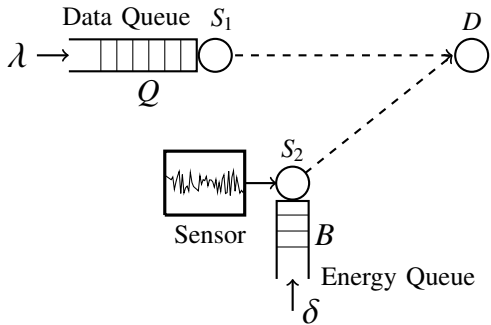
\begin{figure}[h!]
	\vspace{-0.2cm}
	\centering
	\begin{tikzpicture}[scale=0.85]
	\draw[thick, ->] (0,5) -- (0.5,5) ;
	\node[left] at (0,5) {\Large $\lambda$};
	\draw	(1.4,5.9) node[anchor=north] {Data Queue};
	\draw[thick] (0.5,5.25)--(2.5,5.25)-- (2.5,4.75)-- (0.5, 4.75);
	\node[below] at (1.75,4.75) {\large $Q$};
	\draw[-] (2.25,5.25)--(2.25,4.75);
	\draw[-] (2,5.25)--(2,4.75);\draw[-] (1.75,5.25)--(1.75,4.75); \draw[-] (1.5,5.25)--(1.5,4.75); \draw[-] (1.25,5.25)--(1.25,4.75); \draw[-] (1,5.25)--(1,4.75);
	\draw [thick,fill=white](2.79,5)circle[radius= 0.733 em]  ;
	\node[above] at (2.9,5.25) {$S_1$};
	\draw[thick, dashed,->] (3.1,5)--(6.5,5);
	\draw [thick,fill=white](6.8,5)circle[radius= 0.733 em]  ;
		\node[above] at (6.8,5.25) {$D$};
	\draw[thick, dashed,->] (4,3)--(6.5,5);
	\draw [thick,fill=white](4,3)circle[radius= 0.733 em]  ;
	\node[above] at (4,3.2) {$S_2$};
	\draw[thick] (3.75,1.5)--(3.75,2.72)-- (4.25,2.72)-- (4.25, 1.5);
	\draw[-] (3.75,2.5)--(4.25,2.5); 	\draw[-] (3.75,2.25)--(4.25,2.25); 	\draw[-] (3.75,2)--(4.25,2);
	\draw[thick, ->] (4,1) -- (4,1.5) ;
	\node[right] at (4,1) {\Large $\delta$};
	\node[below] at (4.5,2.5) {\large $B$};
	\node[right] at (4.3,1.5) {Energy Queue};
	\draw[thick, ->] (3.25,3) -- (3.75,3) ;
	\draw[very thick] (2,2.5)--(2,3.5)-- (3.25,3.5)-- (3.25, 2.5) --(2,2.5);
	\node[right] at (2,2.2) {Sensor};
	\draw[black] (2,3) \irregularline{0.2cm}{1.25};
	\end{tikzpicture}
		\vspace{-0.15cm}
		\caption{The system model. One throughput-oriented source node and an energy-harvesting (EH) device share the same wireless channel to a common destination. The EH device is generating status updates to transmit to the destination.}
	\label{fig:system}
\end{figure}

We consider a time-slotted MAC where two source nodes with heterogeneous traffic intend to transmit to a common destination $D$, as shown in Fig.~\ref{fig:system}. The first node $S_1$ is connected to the power grid, thus it is not power-limited. Note that $S_1$ is a throughput-oriented node, which intends to achieve as high throughput as possible. The data packets arrives at $S_1$ following a Bernoulli process with probability $\lambda$. We consider an early departure late arrival model for the queue. 
When the data queue of $S_1$ is not empty, it transmits a packet to the destination with probability $q_1$. The second node $S_2$ is not connected to a dedicated power source, but it can harvest energy from its environment, such as wind or solar energy. We assume that the battery charging process follows a Bernoulli process with probability $\delta$, with $B$ denoting the number of energy units in the energy source (battery) at node $S_2$. The capacity of the battery is assumed to be infinite. When $S_2$ has a non-empty battery, it generates a status update with probability $q_2$ and transmits it to the destination.\footnote{A similar update packet generation model for the EH user can be found in \cite{Talak}.}
The transmission of one status update consumes one energy unit from the battery.

We assume multi-packet reception (MPR) capabilities at the destination node $D$, which means that $D$ can decode multiple messages simultaneously with a certain probability. MPR is a generalized form of the packet erasure model, and it captures better the wireless nature of the channel since a packet can be decoded correctly by a receiver that treats interference as noise if the received signal-to-interference-plus-noise ratio (SINR) exceeds a certain threshold. We consider equal-size data packets and the transmission of one packet occupies one timeslot. 

For the notational convenience, we define the following successful transmission/reception probabilities, depending on whether one or both source nodes are transmitting in a given timeslot:
\begin{itemize}
	\item $p_{i/i}$: success probability of $S_i,~i\in\{1,2\}$ when only $S_i$ is transmitting;
	\item $p_{i/i,j}$: success probability of $S_i$ when both $S_i$ and $S_j$ are transmitting;
\end{itemize}

In the case of an unsuccessful transmission from $S_1$, the packet has to be re-transmitted in a future timeslot. We assume that the receiver gives an instantaneous error-free acknowledgment (ACK) feedback of all the packets that were successful in a slot at the end of the
slot. Then, $S_1$ removes the successfully transmitted packets from its buffer. 
In case of an unsuccessful packet transmission from $S_2$, since it contains a previously generated status update, that packet is dropped without waiting to receive an ACK, and a new status update will be generated for its next attempted transmission.

\vspace{-0.2cm}
\subsection{Physical Layer Model}
We consider the success probability of each node $i$ based on the SINR
\begin{equation*}
{\rm SINR}_{i}=\frac{P_{i}|h_{i}|^2\beta_i}{\sum_{j\in \mathcal{A}\backslash\left\{i\right\}} P_{j}|h_{j}|^2 
\beta_j+\sigma^2},
\end{equation*}
where $\mathcal{A}$ denotes the set of active transmitters; $P_{i}$ denotes the transmission power of node $i$; $h_{i}$ denotes the small-scale channel fading from the transmitter $i$ to the destination, which follows $\mathcal{CN}(0,1)$ (Rayleigh fading); $\beta_i$ denotes the large-scale fading coefficient of the link $i$; $\sigma^2$ denotes the thermal noise power. 

Denote by $\theta_i$, $i=\{1,2\}$, the SINR thresholds for having successful transmission. By utilizing the small-scale fading distribution, we can obtain the success probabilities as follows:

\begin{equation}
p_{i/i}=\mathbb{P} \left\lbrace \mathrm{SNR}_{i} \geq \theta_i \right\rbrace = \exp \left(- \frac{\theta_i \sigma^2 }{P_{i}\beta_i}\right),\text{ }i=1,2.
\end{equation}
\vspace{-0.3cm}
\begin{equation} 
p_{i/i,j}=\mathbb{P} \left\lbrace \mathrm{SINR}_{i} \geq \theta_i \right\rbrace = \frac {\exp \left(- \frac{\theta_i \sigma^2 }{P_{i}\beta_i} \right)}{ 1+\theta_i \frac{P_{j}\beta_j}{P_{i}\beta_i} },i=1,2,j\neq i.
\end{equation}

\section{Performance Analysis of Node $S_1$}
\label{sec:primary}
In this section, we study the performance of node $S_1$ regarding (stable) throughput and the average delay per packet needed to reach the destination.
The service probability of $S_1$ is given by
\begin{align}
\mu=&\textnormal{Pr}(B=0)q_1 p_{1/1}+\textnormal{Pr}(B\neq 0)q_1 (1-q_2)p_{1/1}\nonumber\\
&+\textnormal{Pr}(B\neq 0)q_1 q_2p_{1/1,2}\nonumber\\
=&q_1p_{1/1}\left[1-q_2\textnormal{Pr}(B\neq 0)\right]+q_1\textnormal{Pr}(B\neq 0)q_2p_{1/1,2}.
\label{eq:mu}
\end{align}

In this work, we mainly focus on the case where $S_2$ relies on energy harvesting to operate, but for comparison purposes we also consider the case that $S_2$ is connected to the power grid, thus does not have energy limitations.

\subsection{When $S_2$ relies on EH}
When node $S_2$ relies on EH to operate, recall that the energy arrivals at the EH node $S_2$ follows a Bernoulli process with probability $\delta$. The evolution of the energy queue can be modeled as a Discrete Time Markov Chain. Denote by $B$ the energy queue size, we have $\textnormal{Pr}(B\neq 0)=\frac{\delta}{q_2}$ when $\delta<q_2$, which will be the case we consider in the remainder of this paper.\footnote{When $\delta \geq q_2$, the Markov chain is not positive recurrent, thus we do not consider that case.}
Plugging it in \eqref{eq:mu}, we obtain
\begin{equation} \label{eq:mudelta}
\mu=q_1 p_{1/1}(1-\delta)+q_1\delta p_{1/1,2}.
\end{equation}
The queue of $S_1$ is stable if and only if $\lambda<\mu$, which corresponds to $q_1>\frac{\lambda}{p_{1/1}(1-\delta)+\delta p_{1/1,2}}$. When the queue is stable, the probability that the queue is non-empty is $\mathbb{P}[Q\neq 0]=\frac{\lambda}{\mu}$. 

\subsection{When $S_2$ is connected to power grid}
If $S_2$ is connected to the power grid, then we have 
\begin{equation}
\mu=q_1 (1-q_2)p_{1/1}+q_1 q_2p_{1/1,2}.
\end{equation}
The queue is stable if and only if $q_1>\frac{\lambda}{1-q_2p_{1/1}+q_2p_{1/1,2}}$.

The throughput of node $S_1$ is $\mathcal{T}=\min\{\lambda,\mu\}$.
When the queue is stable, the delay of $S_1$, which consists of queueing delay and transmission delay, is given by \cite{srikant2013communication}
\begin{equation}
D=\frac{1-\lambda}{\mu-\lambda}+\frac{1}{\mu}.
\label{eq:delay}
\end{equation}
When the queue is unstable, the delay is infinite due to the infinite queueing delay.

\section{Average AoI of Node $S_2$}

\begin{figure}
	\begin{tikzpicture}[scale=0.95]
	\draw[->] (0,0) -- (8.2,0) node[anchor=north] {$n$};
	\draw[->] (0,0) -- (0,3.5) node[anchor=east] {$\Delta(n)$};
	
	\draw	(-0.3,0.25) node[anchor=south] {$1$};
	\draw	(-0.3,0.7) node[anchor=south] {$2$};
	\draw[thick]  (-0.1,0.5) -- (0.1,0.5); 
	\draw[thick]  (-0.1,1) -- (0.1,1); 
	\draw[thick]  (-0.1,1.5) -- (0.1,1.5); 
	\draw[thick]  (-0.1,2) -- (0.1,2); 
	\draw[thick]  (-0.1,2.5) -- (0.1,2.5); 
	\draw[fill=gray!10] (1,0)--(1,0.5)-- (1.5,0.5)-- (1.5,1)-- (2,1)-- (2,1.5)-- (2.5,1.5)-- (2.5,2)-- (3,2)--(3,0);
	\draw[fill=gray!10] (4,0) --(4,0.5) -- (4.5,0.5) -- (4.5,1)-- (5,1)-- (5,1.5)-- (5.5,1.5)-- (5.5,2)-- (6,2)-- (6,2.5)-- (6.5,2.5)-- (6.5,3)--(7,3)--(7,0);

	\draw[->,>=stealth]    (1,0) -- (1,-0.4) node[anchor=south,below] {$n_1$};
	\draw[->,>=stealth]  (3,0) -- (3,-0.4) node[anchor=south,below] {$n_2$};
	\draw[->,>=stealth]  (4,0) -- (4,-0.4) node[anchor=south,below] {$n_k$};
	\draw[->,>=stealth]   (7,0) -- (7,-0.4) node[anchor=south,below] {$n_{k+1}$};
	
	\draw[<-] (2,0.5) to [out=95,in=250] (1.5,1.5) node [above] {{$Y_1$}};   
	\draw[<-] (5.5,1) to [out=95,in=250] (5,2) node [above] {{$Y_k$}};   
	
	\draw [<->] (4,0.25) -- (5,0.25) node[pos=.75,sloped,above] {$T_1$} ;
	\draw[thick]  (4,0.15) -- (4,0.35);    
	\draw [<->] (6,0.25) -- (7,0.25) node[pos=.5,sloped,above] {$T_M$} ;
	\draw[thick]  (5,0.15) -- (5,0.35);      
	\draw[thick]  (6,0.15) -- (6,0.35);    
	\draw[dotted]  (5,0.25) -- (6,0.25);    
	\draw[thick]  (7,0.15) -- (7,0.35);   
	
	\draw [<->] (1,-1) -- (3,-1) node[pos=.5,sloped,below] {$X_1$} ;
	\draw[thick]  (1,-1.1) -- (1,-0.9) 
	(3,-1.1) -- (3,-0.9);
	
	\draw [<->] (4,-1) -- (7,-1) node[pos=.5,sloped,below] {$X_k$} ;
	\draw[thick]  (4,-1.1) -- (4,-0.9); 
	\draw[thick]  (7,-1.1) -- (7,-0.9) ;
	
	\draw[thick] (0,0.5) -- (0.5,0.5) -- (0.5,1)-- (1,1)-- (1,0.5)-- (1.5,0.5)-- (1.5,1)-- (2,1)-- (2,1.5)-- (2.5,1.5)-- (2.5,2)-- (3,2)--(3,0.5)--(3.5,0.5);
	\draw[white, fill=white!50] (3.5,-0.2) -- (3.5,0.2) -- (3.95,0.2) -- (3.95,-0.2) ;   
	\draw[dashed]  (3.5,0) -- (4,0); 
	\draw[thick] (4,0.5) -- (4.5,0.5) -- (4.5,1)-- (5,1)-- (5,1.5)-- (5.5,1.5)-- (5.5,2)-- (6,2)-- (6,2.5)-- (6.5,2.5)-- (6.5,3)--(7,3)--(7,0.5)--(7.5,0.5);
	
	\draw [decorate,decoration={brace,amplitude=5pt,mirror,raise=2pt},yshift=0pt] (3.1,0.05) -- (3.1,2) node [black,midway,yshift=0.1cm,xshift=0.5cm] {$X_1$};
	\draw [decorate,decoration={brace,amplitude=5pt,mirror,raise=2pt},yshift=0pt] (7.1,0.05) -- (7.1,3) node [black,midway,xshift=0.5cm] { $X_k$};
	\end{tikzpicture}
	\vspace{-0.4cm}
	\caption{Evolution of the AoI. $n_k$ denotes the time when the destination received the $k$-th update. $Y_k$ is the total area below the AoI step line between $n_k$ and $n_{k+1}$. $X_k$ is the number of time slots between two successful receptions of the status updates.}
	\label{fig:aoi}
	\vspace{-0.4cm}
\end{figure}
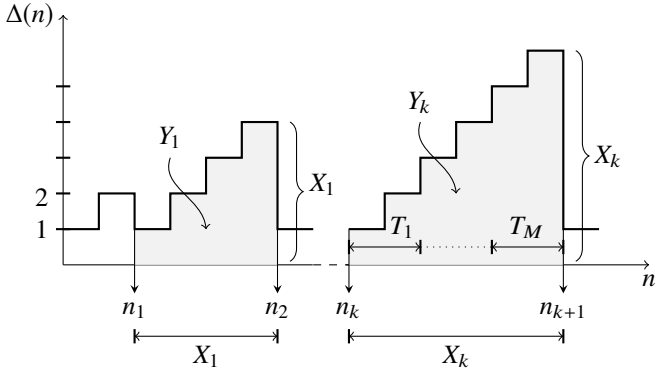

At time slot $n$, the AoI seen at the destination is defined by the difference between the current time $n$ and the generation time $U(n)$ of the latest successfully received update before time $n$, given by
\begin{equation}
\Delta(n)=n-U(n).
\end{equation}

The AoI takes discrete numbers, as shown in Fig.~\ref{fig:aoi}. 

Let $T_{i}$ denote the time between two consecutive attempted transmissions and
$X_k$ denote the waiting time at the destination between the successful reception of the $k$-th and the ($k+1$)-th updates, we have
\begin{equation}
X_k=\sum\limits_{i=1}^{M}T_i,
\label{eq:X_k}
\end{equation} 
where $M$ is a random variable that represents the number of attempted transmissions between two successfully received status updates. 
Note that $X_k$ is a stationary random process, in the following we use $\mathbb{E}[X]$ as the expected value of $X_k$ for an arbitrary $k$.

Recall that one energy chunk is consumed to transmit one status update. For a period of $N$ time slots where $K$ successful updates occur, the average AoI can be computed as 
\begin{equation}
\Delta_N=\frac{1}{N}\sum\limits_{n=1}^{N}\Delta(n)=\frac{1}{N}\sum\limits_{k=1}^{K}Y_k=\frac{K}{N}\frac{1}{K}\sum\limits_{k=1}^{K}Y_k.
\end{equation}
Since $\lim\limits_{N\rightarrow\infty}\frac{K}{N}=\frac{1}{\mathbb{E}[X]}$, and $\frac{1}{K}\sum\limits_{k=1}^{K}Y_k$ is the arithmetic mean of $Y$, which converges almost surely to $\mathbb{E}[Y]$ when $K$ goes to infinity. Then we have
\begin{equation}
\Delta=\lim\limits_{N\rightarrow\infty}\Delta_N=\frac{\mathbb{E}[Y]}{\mathbb{E}[X]}.
\end{equation}
From Fig.~\ref{fig:aoi}, it is not difficult to obtain the relation between $Y_k$ and $X_k$ as follows:
\begin{equation}
Y_k=\sum\limits_{m=1}^{X_k}m=\frac{1}{2}X_{k}(X_{k}+1).
\end{equation}
Then we have
\begin{equation}
\Delta_N=\frac{K}{N}\frac{1}{K}\sum\limits_{k=1}^{K}Y_k=\frac{\mathbb{E}\left[\frac{X_k^2}{2}+\frac{X_k}{2}\right]}{\mathbb{E}[X]}=\frac{\mathbb{E}[X^2]}{2\mathbb{E}[X]}+\frac{1}{2}.
\label{eq:aoi}
\end{equation}

Since $\mathbb{E}[X]$ represents the expected value of $X_k$ for an arbitrary $k$, from \eqref{eq:X_k} we have
\begin{equation} \label{EX-gen}
\mathbb{E}[X]=\sum\limits_{M=1}^{\infty}M \mathbb{E}[T](1-\overline{p}_2)^{M-1}\overline{p}_2=\frac{\mathbb{E}[T]}{\overline{p}_2},
\end{equation}
where $\overline{p}_2$ is the success probability of the transmission from $S_2$, which is the weighted sum of $p_{2/2}$ and $p_{2/1,2}$, given by
\begin{equation}
\overline{p}_2=p_{2/2}(1-q_1 \cdot \text{Pr}[Q\neq 0])+p_{2/1,2} q_1 \text{Pr}[Q\neq 0].
\label{eq:pqne}
\end{equation}
The probability that the information queue of $S_1$ is non-empty depends on the average service rate of $S_1$, which is affected by the activity of the EH sensor because of the interference. The exact expression of $\textnormal{Pr}[Q\neq 0]$ is given in Section~\ref{sec:primary}.

For the second moment of $X$, we start from
\begin{equation}
X_k^2=\left(\sum\limits_{i=1}^{M}T_i\right)^2=\sum\limits_{i=1}^{M}T_i^2+\sum\limits_{i=1}^{M}\sum\limits_{j=1,j\neq i}^{M}T_i T_j.
\end{equation}
Since $T_i$ is a stationary random process, we use $\mathbb{E}[T]$ to represent the expected value of $T_i$ for an arbitrary $i$. Taking conditional expectation of both sides, we get
\begin{equation}
\mathbb{E}[X^2 \vert M]=M\mathbb{E}[T^2]+M(M-1)\left(\mathbb{E}[T]\right)^2.
\end{equation}
Then we have
\begin{align} \label{EX2-gen}
\mathbb{E}[X^2 ]&=\sum\limits_{M=1}^{\infty}\mathbb{E}[X^2 \vert M] (1-\overline{p}_2)^{M-1}\overline{p}_2\nonumber\\
&=\frac{\mathbb{E}[T^2 ]}{\overline{p}_2}+\mathbb{E}[T]^2\frac{2(1-\overline{p}_2)}{\overline{p}_2^2}.
\end{align}
Here, the sum converges when $\overline{p}_2>0$.

After substituting \eqref{EX-gen} and \eqref{EX2-gen} into \eqref{eq:aoi}, we have that the average AoI, $\Delta$, can be written as 
\begin{equation} \label{eq:aoi-et-gen}
\Delta = \frac{\mathbb{E}[T^2 ]}{2\mathbb{E}[T]} + \frac{\mathbb{E}[T](1-\overline{p}_2)}{\overline{p}_2} + \frac{1}{2}.
\end{equation}

Since $T$ represents the time between two consecutive attempted transmissions, we have
\begin{equation}
\begin{split}
\textnormal{Pr}(T=k)=&\textnormal{Pr}(B=0)\sum\limits_{l=1}^{k}(1-\delta)^{k-l}\delta^{l} (1-q_2)^{l-1}q_2\\&+\textnormal{Pr}(B\neq 0)(1-q_2)^{k-1}q_2.
\end{split}
\label{eq:T}
\end{equation}

\subsection{When $S_2$ relies on EH} \label{sec:S2EH}

When $S_2$ relies on harvested energy, recall that we have $\textnormal{Pr}(B\neq 0)=\frac{\delta}{q_2}$ when $\delta<q_2$.
When the queue of $S_1$ is stable, i.e., $\lambda<\mu$, we have $\textnormal{Pr}(Q\neq 0)=\frac{\lambda}{\mu}$ and \eqref{eq:pqne} becomes
\begin{equation}
\overline{p}_2=p_{2/2}-\frac{\lambda(p_{2/2}-p_{2/1,2})}{p_{1/1}(1-\delta)+\delta p_{1/1,2}}.
\label{eq:pavg}
\end{equation}
Note that even though $q_1$ does not appear directly in \eqref{eq:pavg}, it affects the stability condition of the data queue of $S_1$. If the queue is unstable, $\textnormal{Pr}(Q\neq 0)=1$ and \eqref{eq:pqne} becomes $\overline{p}_2=p_{2/2}(1-q_1 )+p_{2/1,2} q_1$.

Let $A=\frac{\delta (1-q_2)}{1-\delta}$, after substituting $\textnormal{Pr}(B\neq 0)=\frac{\delta}{q_2}$ into \eqref{eq:T}, we have
\begin{align}
\textnormal{Pr}(T=k)&= \left(1-\frac{\delta}{q_2}\right)(1-\delta)^k\frac{q_2}{1-q_2}\sum\limits_{l=1}^{k}A^l  +\frac{\delta}{q_2}(1-q_2)^{k-1}q_2  \nonumber\\
&=\left(1-\frac{\delta}{q_2}\right)\frac{q_2(1-\delta)^k}{1-q_2}\frac{A(1-A^k)}{1-A}+\delta(1-q_2)^{k-1}.
\end{align}
Then we obtain
\begin{align}
\mathbb{E}[T]&=\sum\limits_{k=1}^{\infty}k\textnormal{Pr}(T=k)\nonumber \\
&= 
\frac{q_2-\delta}{1-q_2}\frac{A(1-A(1-\delta)^2)(1-\delta)}{\delta^2(1+A(1-\delta)^2)}+\frac{\delta}{q_2}.
\label{eq:ET}
\end{align}
Similarly, we can obtain
\begin{align}
\mathbb{E}[T^2]&=\sum\limits_{k=1}^{\infty}k^2 \textnormal{Pr}(T=k) \nonumber\\
&= \frac{q_2-\delta}{1-q_2}\frac{A}{1-A}\sum\limits_{k=1}^{\infty}k^2(1-\delta)^k(1-A^k)+\frac{\delta(2-q_2)}{q_2^3}.
\label{eq:ETsqure}
\end{align}
Note that the sums in both \eqref{eq:ET} and \eqref{eq:ETsqure} converge when $\delta>0$.

The summation $\sum\limits_{k=1}^{\infty}k^2(1-\delta)^k(1-A^k)$ has a closed form expression which is lengthy and provides little insights. Thus, it is omitted here for neater presentation of the results.

\subsection{When $S_2$ is connected to a power grid}
In order to show the effect of EH on the AoI, here we consider the case where $S_2$ is instead connected to a power grid without the need to harvest ambient energy, which is also a baseline case for the considered network.
Since $\textnormal{Pr}(B\neq 0)=1$, from \eqref{eq:T} we have
	\begin{equation}
	\textnormal{Pr}(T=k)=(1-q_2)^{k-1}q_2.
	\end{equation}
Then we obtain
\begin{eqnarray}
\mathbb{E}[T]&=&\sum\limits_{k=1}^{\infty}k\textnormal{Pr}(T=k)=\frac{1}{q_2},\\
\mathbb{E}[T^2]&=&\sum\limits_{k=1}^{\infty}k^2 \textnormal{Pr}(T=k)=\frac{2-q_2 }{q_2^2}.
\end{eqnarray}
After plugging them into \eqref{eq:aoi-et-gen}, we obtain
\begin{equation}
\Delta=\frac{1}{q_2 \overline{p}_2},
\end{equation}
where $\overline{p}_2=p_{2/2}-\frac{\lambda(p_{2/2}-p_{2/1,2})}{p_{1/1}(1-q_2)+ q_2 p_{1/1,2}}$, when the queue at $S_1$ is stable, otherwise $\overline{p}_2=p_{2/2}(1-q_1 )+p_{2/1,2} q_1$.

\section{Numerical Evaluation}
In this section, we evaluate the delay of node $S_1$ and the average AoI of node $S_2$ to illustrate the relation between these two metrics and the impact of different system parameters on these two performance metrics. 

Since our analytical results of the delay and AoI do not require any specific channel model, the parameters we use in this section are: $\frac{P_1\beta_1}{\sigma^2}=11$\,dB, $\frac{P_2\beta_2}{\sigma^2}=13$\,dB.

 \begin{figure}[b]
	\centering
	\includegraphics[width=0.9\columnwidth]{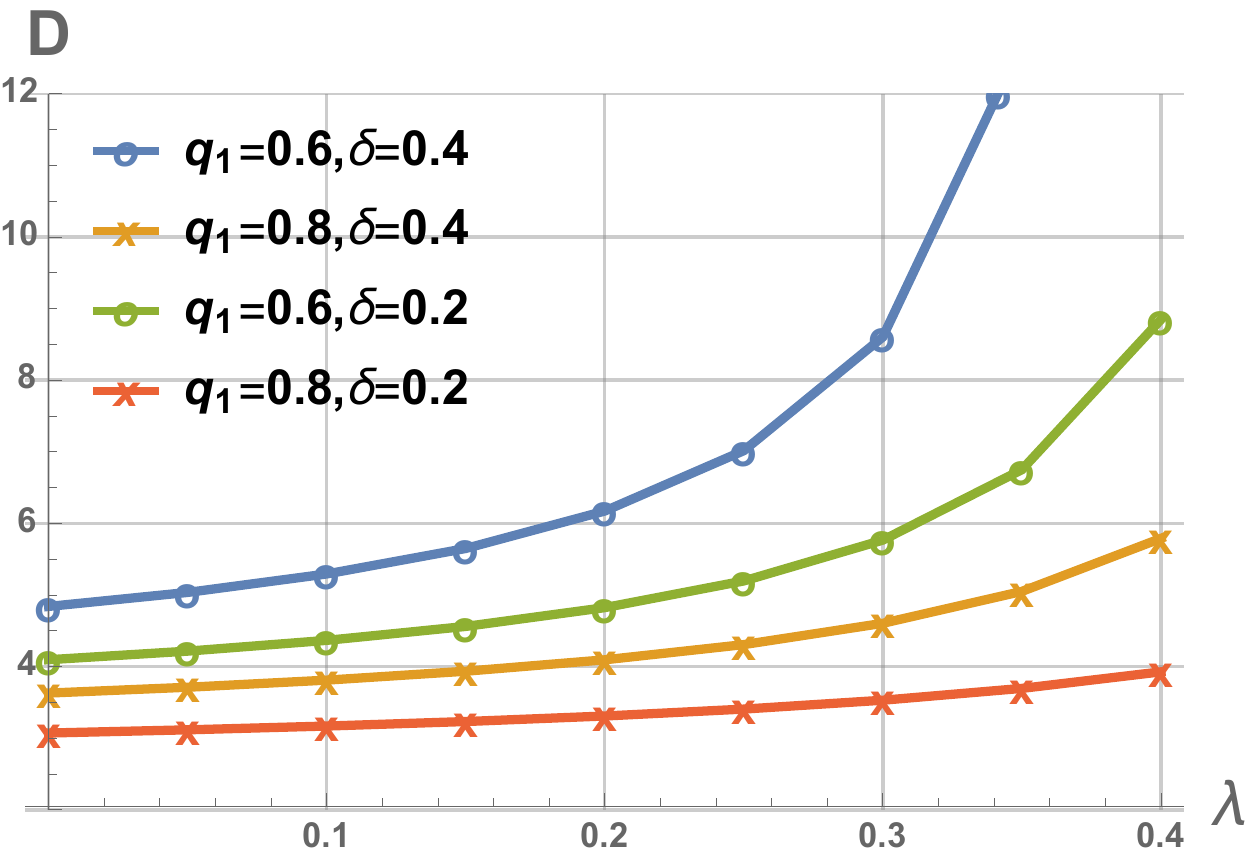}
	\caption{Delay of $S_1$ as a function of the data arrival rate $\lambda$. The transmission probability of $S_2$ is $q_2=0.5$. $\theta_1=\theta_2=\theta=0$\,dB.}
	\label{fig:delay_vs_lambda}
\end{figure}

\begin{figure}[h!]
	\centering
	\includegraphics[width=0.9\columnwidth]{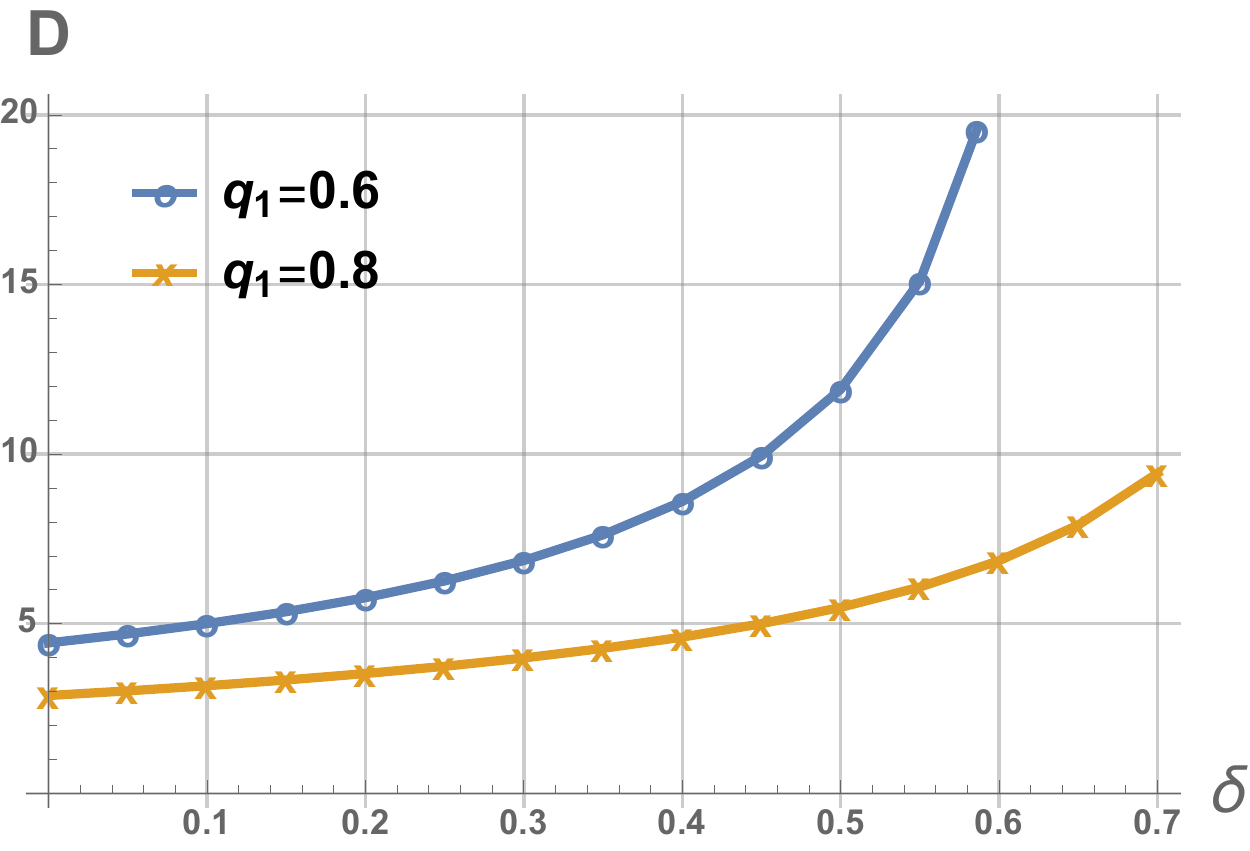}
	\caption{Delay of $S_1$ as a function of the energy charging rate $\delta$. $q_2=0.8$, $\lambda=0.3$. $\theta_1=\theta_2=\theta=0$\,dB.}
	\label{fig:delay_vs_delta}
\end{figure}

\subsection{Delay of Node $S_1$}
In Figs.~\ref{fig:delay_vs_lambda} and \ref{fig:delay_vs_delta}, we plot the delay per packet for node $S_1$ as functions of the data arrival rate $\lambda$ and the energy arrival rate $\delta$ of the EH node, respectively. It is obvious that the delay increases with $\lambda$ and $q_2$, as we can see from \eqref{eq:delay}. Note that for Fig.~\ref{fig:delay_vs_lambda}  we only present the values of $D$ for $\lambda\in[0, 0.4]$, because when $\lambda$ is too large, the queue of $S_1$ becomes unstable and the delay is infinite.

\subsection{AoI of Node $S_2$}
\begin{figure}[h!]
	\centering
	\includegraphics[width=0.9\columnwidth]{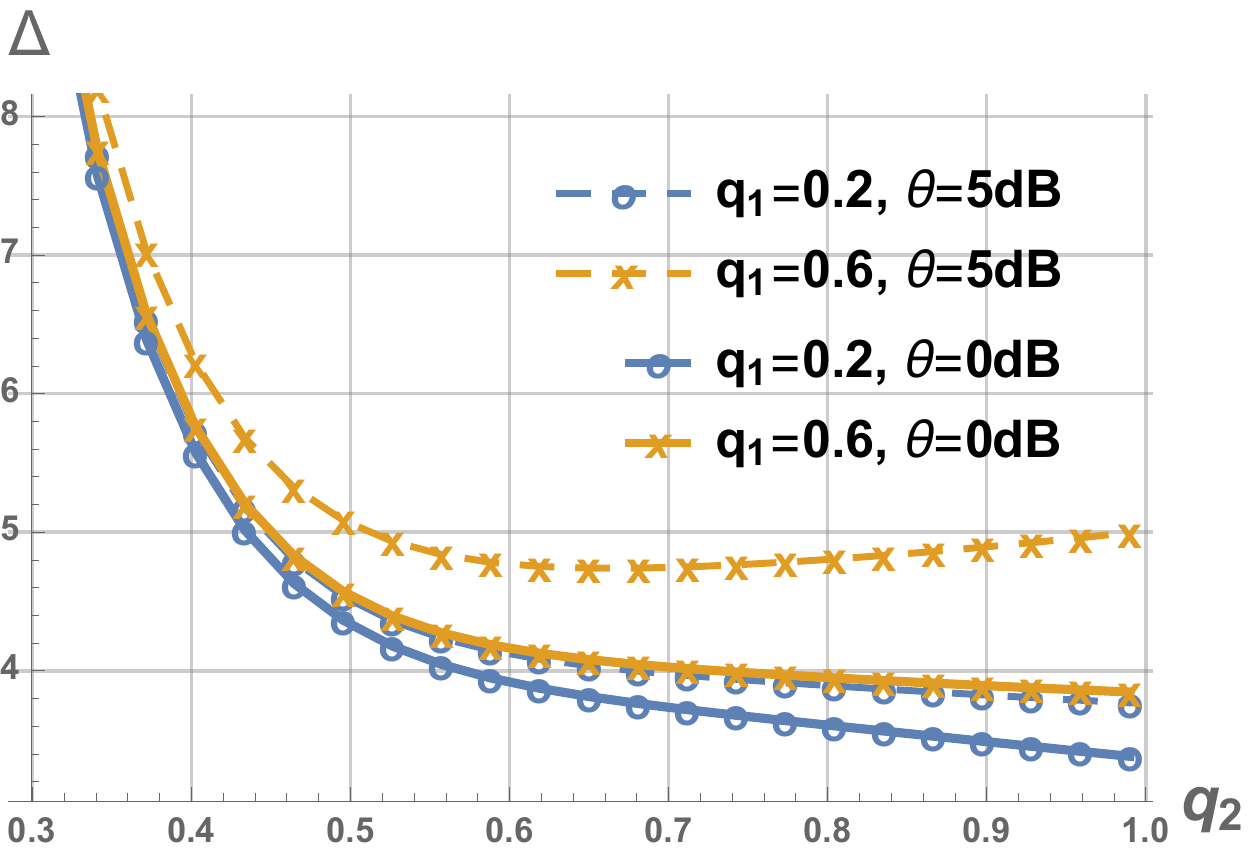}
	\caption{AoI of the EH node $S_2$ vs. transmission probability $q_2$.  $\lambda=0.6$, $\delta=0.3$.}
	\label{fig:aoi_vs_q2_eh}
\end{figure}

\begin{figure}[ht!]
	\centering
	\includegraphics[width=0.9\columnwidth]{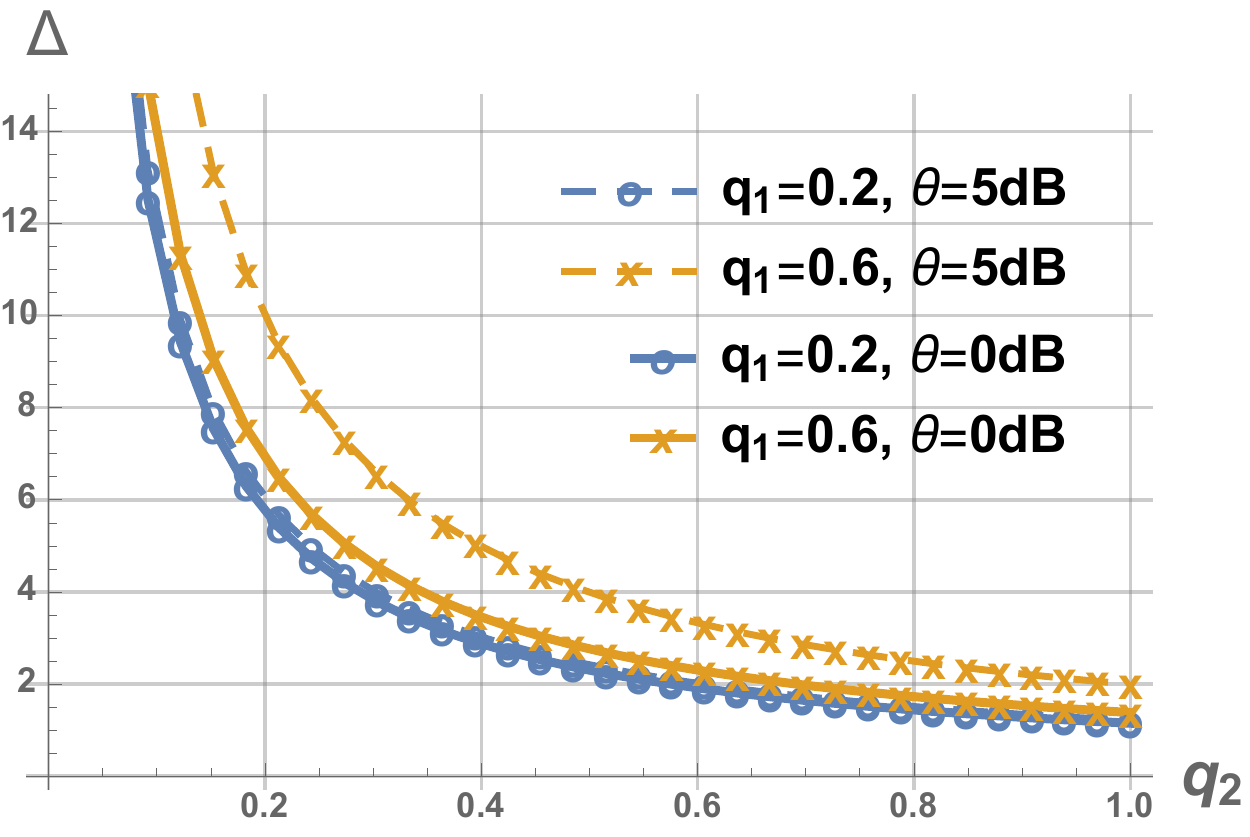}
	\caption{AoI of node $S_2$ vs. $q_2$, when $S_2$ is connected to the power grid.  $\lambda=0.6$.}
	\label{fig:aoi_vs_q2}
\end{figure}

In Figs. \ref{fig:aoi_vs_q2_eh} and \ref{fig:aoi_vs_q2}, we plot the AoI of the node $S_2$ as a function of the transmission probability $q_2$, for the cases when $S_2$ relies on EH and when it is connected to a power grid, respectively.
First, from both figures we observe that when $\theta$ is higher, the AoI is larger. This is expected since a higher SINR threshold gives a lower transmission success probability, which consequently increases the AoI.
Second, the most noticeable difference between these two figures is that, in Fig.~\ref{fig:aoi_vs_q2_eh} for the case with $q_1=0.6$ and $\theta=5$\,dB, the AoI first decreases then increases with $q_2$, while in the other cases AoI monotonically decreases with $q_2$. This is because the AoI can be interference-limited or battery-limited depending on whether it relies on EH or not. In Fig.~\ref{fig:aoi_vs_q2_eh}, $S_2$ only relies on harvested energy and the energy arrival rate is relatively small ($\delta=0.3$). When $q_2$ increases to the a certain level, keep increasing the transmission probability $q_2$ might not help reducing the AoI because the collision probability increases, and the energy chunks might be wasted when $S_2$ attempts to transmit too often.

\begin{figure}[ht!]
	\centering
	\includegraphics[width=0.9\columnwidth]{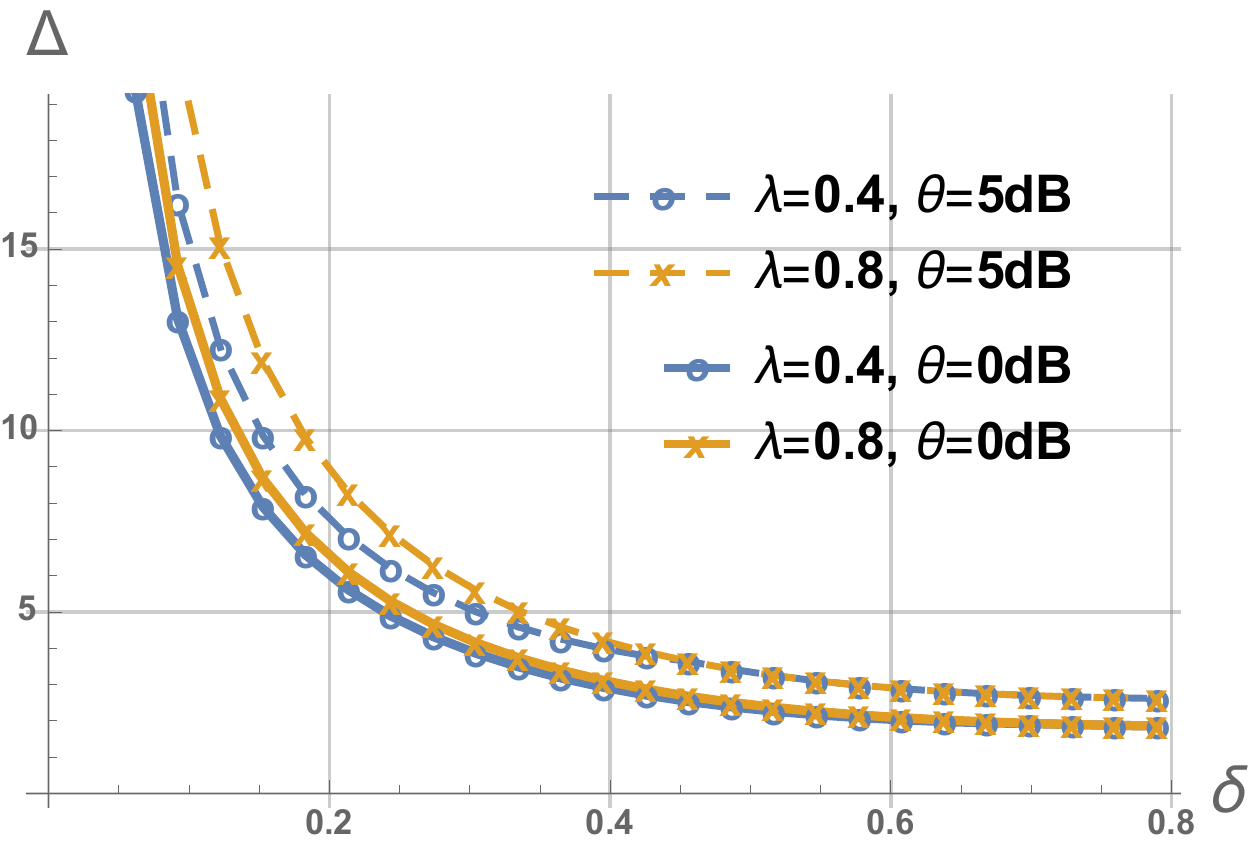}
	\caption{AoI of the EH node $S_2$ vs. energy arrival rate $\delta$. $q_1=q_2=0.8$. }
	\label{fig:aoi_vs_delta_eh}
\end{figure}

In Fig.~\ref{fig:aoi_vs_delta_eh}, we show the AoI of $S_2$ as a function of the energy arrival rate $\delta$. 
In all cases, the AoI decreases with $\delta$, since $\delta$ increases the probability that $S_2$ is charged. 

\begin{figure}[t]
	\centering
	\includegraphics[width=0.9\columnwidth]{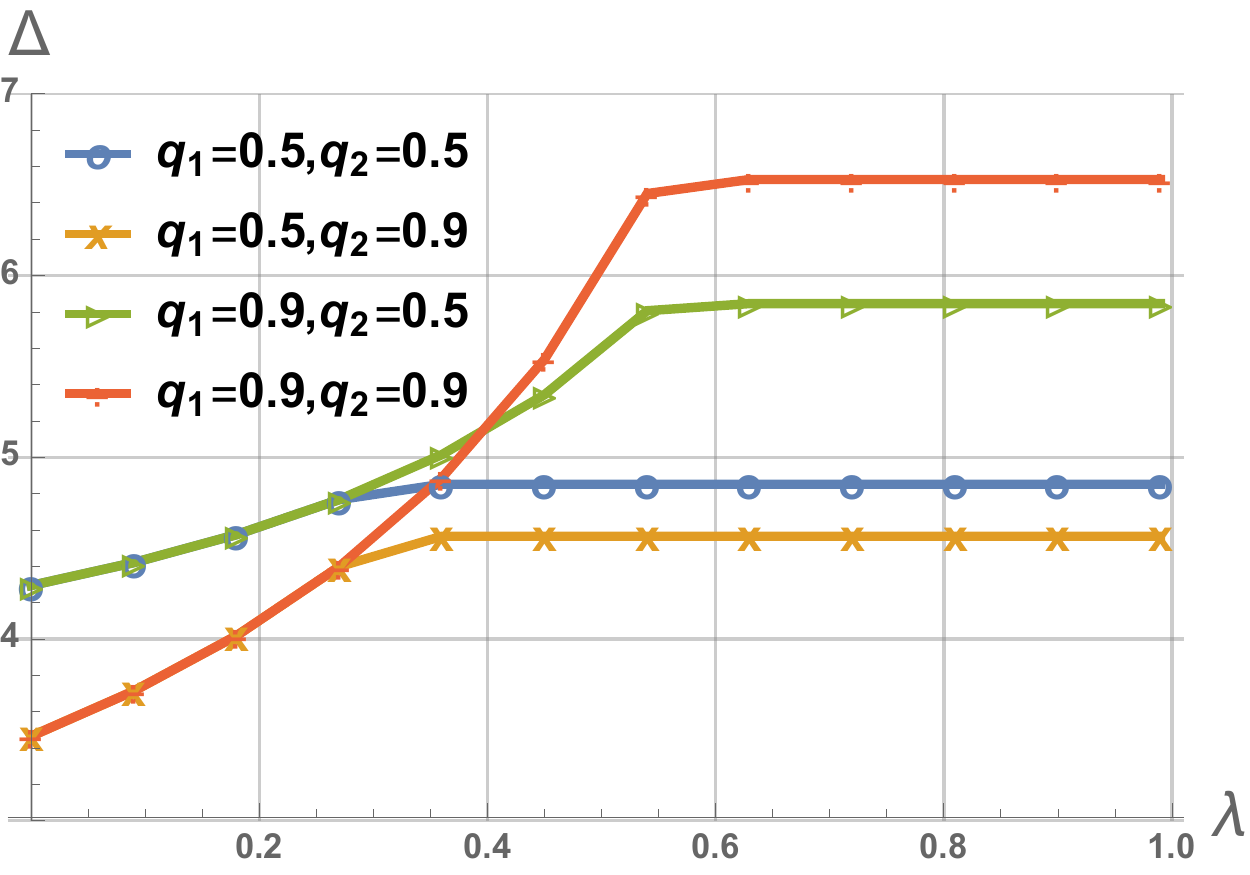}
	\caption{AoI of the EH node $S_2$ vs. data arrival rate $\lambda$ of node $S_1$. The energy charging rate of the EH node is $\delta=0.3$. $\theta_1=\theta_2=\theta=5$\,dB.}
	\label{fig:aoi_vs_lambda_eh}
\end{figure}

\begin{figure}[t]
\centering
\includegraphics[width=0.9\columnwidth]{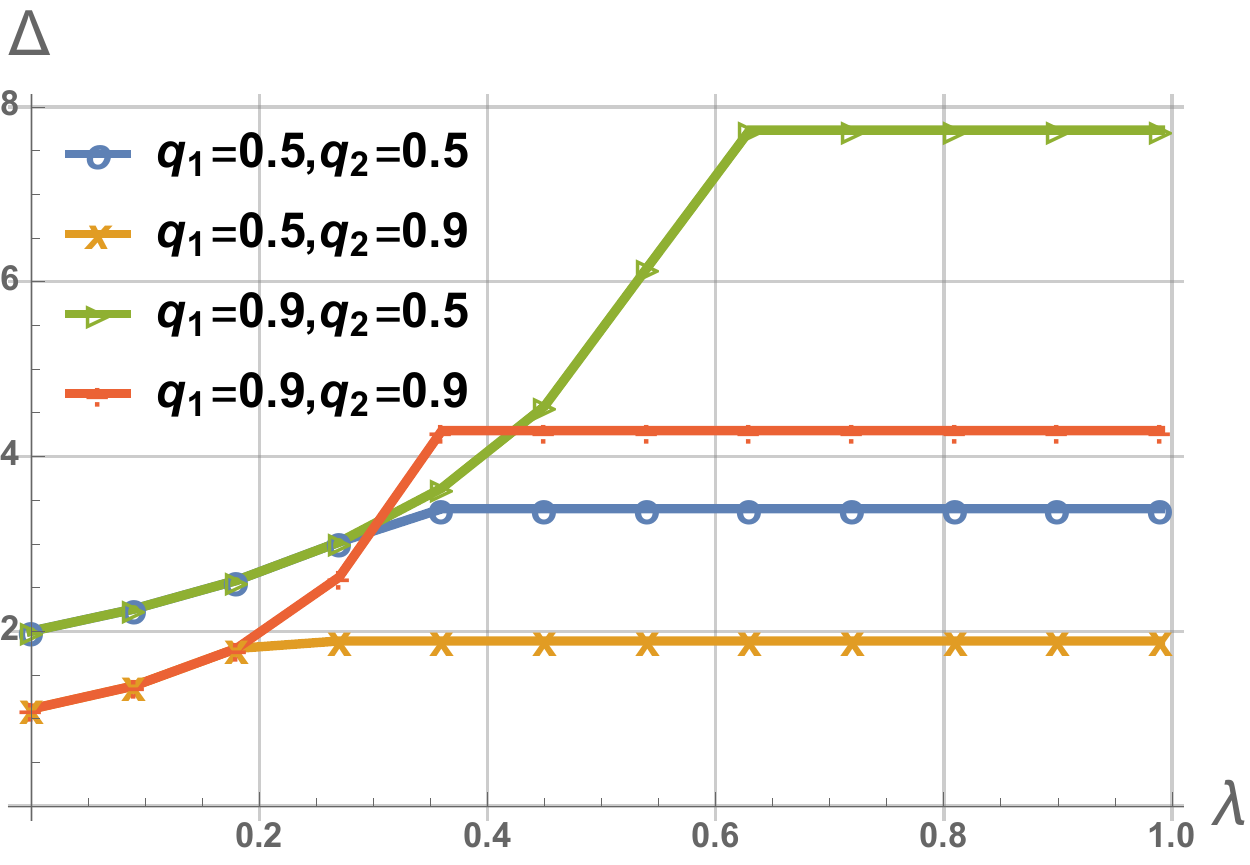}
\caption{AoI of node $S_2$ vs. $\lambda$, when $S_2$ is connected to the power grid. $\theta_1=\theta_2=\theta=5$\,dB.}
\label{fig:aoi_vs_lambda}
\end{figure}

In Figs. \ref{fig:aoi_vs_lambda_eh} and \ref{fig:aoi_vs_lambda}, we plot the AoI of node $S_2$ as a function of the data arrival rate $\lambda$ of node $S_1$, for the cases when $S_2$ relies on EH and when it is connected to the power grid, respectively.
For all the cases, the AoI first increases with $\lambda$, and then saturates. This is because the throughput of node $S_1$ is $\min\{\lambda,\mu\}$. When the throughput is limited by the arrival rate $\lambda$, increasing $\lambda$ means higher probability $\mathbb{P}[Q\neq 0]$, thus higher interference to the transmission of status updates from $S_2$. When the throughput is limited by the service rate $\mu$, increasing $\lambda$ will not change the interference because the queue of $S_1$ is always saturated, and the AoI will remain the same.

\section{Conclusions}
In this work, we studied the performance of a multiple access channel with heterogeneous traffic: one grid-connected node has bursty data arrivals and another node with energy harvesting capabilities sends status updates to a common destination. We derived closed-form expressions for the delay of the first node and the age of information of the second node, which depend on several system parameters such as the transmission probabilities, the data and energy arrival rates. Our results provide fundamental understanding of the delay and age performance and tradeoffs in interference-limited networks with heterogeneous nodes.

\bibliographystyle{IEEEtran}
\bibliography{ref.bib}
\end{document}